# Front-end and Back-end Computational Modeling of 40-Hz Auditory Steady-State Response Abnormalities in Schizophrenia

Wenjun Xia1*, Yan Xu1 and Zhengdi Zhang1*

1School of Mathematical Sciences, Jiangsu University, Zhenjiang, China.

*Corresponding author(s). E-mail(s): wjqx@ujs.edu.cn(first author); dyzhang@ujs.edu.cn.

## Abstract

**Background and Hypothesis.** The 40-Hz auditory steady-state response (ASSR) is frequently reduced in schizophrenia, but group-level electroencephalography (EEG) measures do not identify whether this contrast reflects altered auditory input transformation, cortical excitation/inhibition (E/I) dynamics, or both. We hypothesized that similar ASSR group differences could be reproduced by distinct model-level mechanisms.

**Study Design.** EEG-derived gamma-band suprathreshold proportion (γ%) and 40-Hz inter-trial phase consistency (ITPC) from 21 healthy controls (HC) and 21 participants with schizophrenia (SCZ) constrained a signal-processing auditory front-end coupled to a Wilson–Cowan E/I model. We compared front-end-restricted, back-end-restricted, and full-joint parameter searches, followed by local parameter perturbation and fixed-point analyses.

**Study Results.** HC means exceeded SCZ means for both γ% and ITPC, although participant-level group comparisons were not statistically significant. All three model configurations satisfied the prespecified fitting criteria and reproduced the HC > SCZ direction for both primary EEG-derived measures, while locating group differences in different model components. The front-end-restricted model reproduced the phenotype through differences in effective auditory input transformation, the back-end-restricted model through differences in Wilson–Cowan cortical dynamics, and the full-joint model through combined changes across both stages. Local perturbation analysis showed configuration-dependent robustness, with the full-joint solution exhibiting the highest tested joint γ%–ITPC direction-preservation rates. Fixed-point analysis further showed that similar ASSR outputs could coexist with distinct local dynamical organizations.

**Conclusions.** All three models reproduced the characteristic HC > SCZ direction reported in schizophrenia-related 40-Hz ASSR studies, and suggest that schizophrenia-related neurophysiological abnormalities may reflect distinct pathophysiological pathways, including altered auditory sensory encoding/effective input transformation, altered cortical E/I dynamics, or combined alterations across both stages. This three-model framework provides a computational basis for future patient-level mechanistic comparison, and following clinical validation, may support individualized mechanistic assessment and clinical stratification.

## Introduction

The auditory steady-state response (ASSR) is a frequency-locked neural response elicited by periodic or amplitude-modulated auditory stimulation.[1,2] The 40-Hz ASSR lies within the gamma-frequency range and is commonly associated with temporal locking and synchronized activity across auditory neural populations.[1,3,4] Previous EEG and MEG studies, together with meta-analyses, have repeatedly reported reduced 40-Hz ASSR power and phase locking in schizophrenia, supporting its use as a candidate electrophysiological marker of circuit dysfunction.[5-12]

Mechanistically, gamma-band synchronization and 40-Hz generation have been linked to parvalbumin-positive GABAergic interneurons, NMDA receptor function, and cortical excitation/inhibition (E/I) interactions.[3,13-18] The Wilson-Cowan model describes the nonlinear dynamics of interacting excitatory and inhibitory neural populations and provides a classical framework for studying E/I interactions, fixed-point stability, oscillatory responses, and input-dependent dynamics.[19] Previous computational studies have further shown that changes in inhibitory kinetics or parvalbumin-interneuron-related circuitry can reshape gamma-range auditory entrainment in schizophrenia models.[20,21] Combining 40-Hz ASSR phenotypes with a Wilson--Cowan model may therefore help examine potential dynamical mechanisms underlying gamma synchrony abnormalities in schizophrenia.

Reduced ASSR, however, need not arise solely from cortical E/I dynamics. Before cortical processing, auditory stimulation undergoes frequency-selective filtering, nonlinear compression, and temporal integration,[1,22,23] and ASSR generators span both cortical and subcortical structure.[4] Different changes in effective auditory input or cortical dynamics may therefore produce similar macroscopic EEG phenotypes.

We developed a model combining an auditory front-end with a Wilson-Cowan E/I back-end and constrained it with EEG-derived gamma-band suprathreshold proportion ($\gamma$%) and 40-Hz inter-trial phase consistency (ITPC). We compared front-end-restricted, back-end-restricted, and full-joint parameter searches, followed by local perturbation and dynamical analyses. The aim was not to assign a single pathological cause, but to test whether distinct model-level mechanisms could reproduce the same HC-SCZ response direction under common empirical constraints.

# Methods

## EEG data, empirical comparisons, and model targets

The 40-Hz ASSR recordings used as the empirical basis for model constraints were drawn from the public ASZED schizophrenia EEG dataset.[24,25] To reduce heterogeneity associated with the auditory context, the analysis was restricted to recordings acquired under a single common language condition. Only participants with complete 40-Hz ASSR EEG data under this condition were retained, yielding a final EDF-based subset of 42 participants: 21 healthy controls (HC) and 21 participants with schizophrenia (SCZ). The present study used de-identified, publicly available ASZED data. According to the original dataset report, data collection was approved by the Obafemi Awolowo University Teaching Hospital College Ethics Committee, was conducted in accordance with the Declaration of Helsinki, and included written informed consent from participants or their legal guardians.[3224] For each participant, the available EEG channels were averaged to obtain a single time series. Two subject-level measures were then extracted using the same signal definitions applied to the model output: gamma-band suprathreshold proportion (γ%) and 40-Hz inter-trial phase consistency (ITPC).

Let $\gamma_{g,j}$ and $c_{g,j}$ denote the γ% and ITPC values for participants $j$ in group $g \in \{\mathrm{HC}, \mathrm{SCZ}\}$. Groups means were

$$\overline{\gamma}_g = \frac{1}{N_g} \sum_{j=1}^{N_g} \gamma_{g,j}, \qquad \overline{c}_g = \frac{1}{N_g} \sum_{j=1}^{N_g} c_{g,j} \tag{1}$$

The EEG-derived values used as model-fitting constraints were

$$\overline{\gamma}_{HC} = 41.496, \quad \overline{\gamma}_{SCZ} = 35.476, \tag{2}$$

$$\overline{c}_{HC} = 0.325, \quad \overline{c}_{SCZ} = 0.258, \tag{3}$$

so that

$$y_{target} = (41.496, 35.476, 0.325, 0.258). \tag{4}$$

Participant-level group contrasts were characterized separately from model fitting. Welch two-sample *t* tests were used as the primary parametric comparisons. For each HC-SCZ mean difference, a 95% percentile confidence interval was obtained from 10000 independent bootstrap resamples within the two groups, and Hedges' $g$ was reported as a standardized effect size. Two-sided Mann-Whitney tests were used as distribution-robust sensitivity analyses because normality tests indicated departures from normality for at least one group. Because γ% and ITPC constituted two simultaneous empirical comparisons, the two Welch *P* values were additionally adjusted by the Holm procedure. These inferential analyses characterized uncertainty in the empirical contrasts; the modeling analyses used the prespecified group means above as numerical constraints and did not treat statistical significance as a fitting criterion.

## ASSR measures

In this study, γ% does not refer to conventional gamma power. Instead, it denotes the proportion of post-baseline samples at which the 24-64 Hz envelope exceeds a baseline-derived threshold. This measure therefore quantifies the temporal proportion of suprathreshold gamma-band activity after the baseline interval.

Let $x[n]$ denote either an EEG signal or a model output signal sampled at frequency $f_s$, and let $B_\gamma[\cdot]$ denote the 24-64 Hz band-pass operation. The filtered signal was

$$x_\gamma[n] = B_\gamma[x[n]]. \tag{5}$$

Let $\mathcal{H}\{\cdot\}$ denote the ordinary Hilbert transform. The corresponding analytic signal and envelope were

$$z_\gamma[n] = x_\gamma[n] + i\,\mathcal{H}\{x_\gamma[n]\}, \tag{6}$$

$$a_\gamma[n] = |z_\gamma[n]|. \tag{7}$$

The gamma-band signal was obtained using a fourth-order Butterworth band-pass filter with zero-phase forward-backward filtering. Let $N_{base} = \lfloor 0.5 f_s \rfloor$ denote the number of samples in the 0.5-s baseline. The baseline mean and population standard deviation were

$$\mu_{base} = \frac{1}{N_{base}} \sum_{n=0}^{N_{base}-1} a_\gamma[n], \tag{8}$$

$$\sigma_{base} = \sqrt{\frac{1}{N_{base}} \sum_{n=0}^{N_{base}-1} (a_\gamma[n] - \mu_{base})^2}. \tag{9}$$

The denominator $N_{base}$ is consistent with the default population standard deviation used by *numpy.std*. The baseline threshold was

$$\theta_\gamma = \mu_{base} + 2\sigma_{base} . \quad (10)$$

For a signal containing $N$ samples, let $N_{post} = N - N_{base}$ . The gamma suprathreshold proportion was

$$\gamma\% = \frac{100}{N_{base}} \sum_{n=N_{base}}^{N-1} \mathbf{1}\left[a_\gamma[n] > \theta_\gamma\right] , \quad (11)$$

where $\mathbf{1}[\cdot]$ is the indicator function. A larger $\gamma\%$ indicates that the gamma-band envelope remained above the baseline-derived threshold for a larger proportion of the post-baseline recording.

ITPC was used to quantify the phase consistency of 40-Hz narrowband activity across trials or epochs, following the general phase-synchrony framework based on unit phase vectors[26]. Let $x_k[n]$ denote the signal from trial $k$, and let $B_{40}[\cdot]$ denote the 38-42 Hz band-pass operation. The filtered signal and instantaneous phase were

$$x_{40,k}[n] = B_{40}[x_k[n]] , \quad (12)$$

$$\phi_k[n] = arg\left[x_{40,k}[n] + i\,\mathcal{H}\left\{x_{40,k}[n]\right\}\right]. \quad (13)$$

At each sample $n$, ITPC was defined as the magnitude of the average unit phase vector:

$$\mathrm{ITPC}[n] = \left|\frac{1}{K} \sum_{k=1}^{K} e^{i\phi_k[\mathrm{n}]}\right|, \quad (14)$$

where $K$ is the number of trials or epochs. The final ITPC value was obtained by averaging over the analysis index set $\mathcal{A}$:

$$\mathrm{ITPC} = \frac{1}{|\mathcal{A}|} \sum_{\mathrm{n} \in \mathcal{A}} \mathrm{ITPC}[n]. \quad (15)$$

ITPC was computed using a fourth-order 38-42 Hz Butterworth band-pass filter with zero-phase forward-backward filtering. For the model analysis, each 2.0-s stimulus trial used an analysis window $\mathcal{A}$ spanning 0.25-1.75 s relative to stimulus onset, so that both the first and final 0.25 s were excluded. For the empirical EEG analysis, the signal after the initial 0.5-s baseline was divided into non-overlapping 2.0-s analysis epochs, and the same 0.25-1.75 s within-epoch window was used. Thus, empirical ITPC was treated as an epoch-based 40-Hz phase-consistency measure under this preprocessing scheme, whereas model ITPC was computed across separately simulated stimulus trials. The same Hilbert phase procedure was applied to the EEG signals and to the model output $E(t)$, implemented as $\tau_E$ in the simulation code.

## Auditory front-end

The auditory front-end transformed the 40-Hz amplitude-modulated sound into an effective input $I_{ext}(t)$ that drives the Wilson-Cowan back-end. This module was not intended to fully reproduce the peripheral auditory pathway. Instead, it provided a simplified signal-processing approximation motivated by common stages in auditory models, including frequency-selective filtering, nonlinear transduction, and temporal integration[22,23]. The lateral-inhibition-like channel-difference operation used here was an effective modeling step rather than a direct implementation of a specific anatomical circuit.

Each simulated trial contained a 0.5-s silent baseline, a 2.0-s stimulus interval, and a 0.5-s silent post-stimulus interval. The audio sampling rate was 16 kHz, the carrier frequency was $f_c = 1000\text{Hz}$, the modulation frequency was $f_m = 40\text{Hz}$, and a 10-ms cosine ramp was applied at stimulus onset and offset. For trial $k$ in group $g$, the stimulus was

$$s_{g,k}(t) = r(t)\left[\frac{1+m\sin(2\pi f_m t + \phi_{g,k})}{2}\right]\sin(2\pi f_c t)\,, \tag{16}$$

where $m$=0.5771515 was the fixed modulation depth and $r(t)$ was the onset/offset ramp. The trial-wise phase perturbation was sampled as

$$\phi_{g,k} \sim \mathcal{N}(0,\sigma^2_{\phi,g}) \,, \tag{17}$$

with fixed group-specific dispersions

$$\sigma_{\phi,\mathrm{HC}} = 1.34984,\ \sigma_{\phi,\mathrm{SCZ}} = 1.79279 \,. \tag{18}$$

These phase-dispersion settings were held fixed in all three experiments and were not included among the independently searched front-end parameters.

The front-end first applied a bank of constant- $Q$ band-pass filters with logarithmically spaced center frequencies. Let $f_q$ denote the center frequency of channel $q$, with $n_{\mathrm{ch}}$ total channels. Then

$$f_q = f_{min} \left( \frac{f_{\max}}{f_{\min}} \right)^{\frac{q-1}{n_{\mathrm{ch}}-1}} , \quad q=1,2,\ldots,n_{\mathrm{ch}} \,. \tag{19}$$

The upper frequency was determined by

$$f_{max} = f_{min} 2^{n_{oct}} \,, \tag{20}$$

where $f_{\min}$ is the lowest center frequency and $n_{\mathrm{oct}}$ is the number of octaves covered by the filterbank. The approximate bandwidth of channel $q$ was

$$\Delta f_q = \frac{f_q}{Q_{ERB}} \,. \tag{21}$$

The filter edges were constrained to the valid audio range:

$$f_{q,low} = max(f_q - \frac{\Delta f_q}{2}, 10) \,, \tag{22}$$

$$f_{q,high} = min(f_q + \frac{\Delta f_q}{2}, \frac{f_s}{2} - 10) \,. \tag{23}$$

For the final full-joint search and the formal sensitivity analysis, an additional digital-filter feasibility condition was imposed so that the highest constant-$Q$ band did not rely on this upper-edge truncation:

$$f_{max}\,(1+\frac{1}{2Q_{ERB}}) \leq \frac{f_s}{2}\text{-10 Hz} . \quad (24)$$

Candidates or perturbed front-end realizations violating Eq.24 were rejected and redrawn. All final accepted front-end parameterizations satisfied this condition.

Each channel was implemented as a fourth-order Butterworth band-pass filter with zero-phase forward-backward filtering. Its output was denoted as

$$y_q(t) = B_q[s(t)]. \quad (25)$$

A hair-cell-like nonlinear transformation was then applied. The temporal derivative was

$$d_q(t) = \frac{dy_q(t)}{dt}\,, \quad (26)$$

and the signed power-law compression was

$$h_q(t) = sgn\,(d_q(t))\left|d_q(t)\right|^g . \quad (27)$$

The compressed output was filtered using a first-order Butterworth low-pass filter:

$$\tilde{h}_q(t) = L_{f_{LP}}[h_q(t)] . \quad (28)$$

To approximate lateral-inhibition-like interactions across frequency channels, adjacent-channel differences were rectified:

$$u_q(t) = max\,(\tilde{h}_q(t) - \tilde{h}_{q-1}(t), 0) . \quad (29)$$

A first-order temporal integration stage was used to represent midbrain-like temporal smoothing:

$$\tau_{MB} \frac{dv_q(t)}{dt} = -v_q(t) + u_q(t)\ . \tag{30}$$

In the numerical implementation, this stage was evaluated using the causal first-order IIR recursion

$$v_q[n] = \alpha v_q[n\text{-}1] + (1\text{-}\alpha)\ u_\mathrm{q}[n]\ , \tag{31}$$

where

$$\alpha = exp\ (-\frac{\Delta t_{audio}}{\tau_{MB}})\ . \tag{32}$$

The integrated outputs were averaged across channels and min-max normalized:

$$\tilde{I}_{ext}\ [n] = \frac{1}{n_\mathrm{ch}} \sum_{q=1}^{n_\mathrm{ch}} v_q\ [n]\ , \tag{33}$$

$$I_{ext}[n] = \frac{\tilde{I}_{ext}\ [n] - min_n\ \tilde{I}_{ext}\ [n]}{max_n \tilde{I}_{ext}\ [n] - min_n\ \tilde{I}_{ext}\ [n] + \varepsilon}\ , \tag{34}$$

where $\varepsilon = 10^{-12}$ was used to avoid division by zero. The normalized signal was resampled from 16 kHz to 1 kHz before entering the Wilson-Cowan back-end.

Thus, the auditory front-end can be summarized as

$$I_{ext}(t) = F_{FE}(s(t)\ ; Q_{ERB}, f_{min}, n_{oct}, f_{LP}, g, \tau_{MB})\ . \tag{35}$$

The independently searched front-end parameter set was

$$\Theta_{FE} = \{Q_{ERB}, f_{min}, n_{oct}, f_{LP}, g, \tau_{MB}\}\ . \tag{36}$$

The number of channels was fixed at $n_\mathrm{ch} = 128$ in the principal simulations. The parameter $f_\mathrm{max}$ was derived from $f_\mathrm{min}$ and $n_\mathrm{oct}$ and was not treated as an independent search parameter.

## Wilson-Cowan back-end

The cortical back-end was modeled using a Wilson-Cowan E/I neural population system.[19] The model describes the mean activity of an excitatory population, denoted by $E(t)$, and an inhibitory population, denoted by $I(t)$. The auditory front-end output $I_{\text{ext}}(t)$ was used as the external drive to the excitatory population. The continuous-time model equations were

$$\tau_E \frac{dE(t)}{dt} = -E(t) + S(w_{EE}\, E(t) - w_{EI}\, I(t) + I_{ext}(t) + \eta_E(t)) , \quad (37)$$

$$\tau_I \frac{dI(t)}{dt} = -I(t) + S(w_{IE}\, E(t) - w_{II}\, I(t) + \eta_I(t)) , \quad (38)$$

where $\tau_E$ and $\tau_I$ are the excitatory and inhibitory time constants, respectively, and $w_{EE}$, $w_{EI}$, $w_{IE}$, and $w_{II}$ are effective coupling parameters. The activation function was a logistic sigmoid,

$$S(u) = \frac{1}{1 + \exp[\, -(u-\theta)/\sigma\, ]} , \quad (39)$$

where $\theta$ controls the response threshold and $\sigma$ controls the slope of the nonlinear input-output function.

The noise terms were implemented as independent Gaussian input samples generated at each numerical time step:

$$\eta_E[n] = \sigma_\eta\, \xi_E[n], \quad \eta_I[n] = \sigma_\eta\, \xi_I[n] , \quad (40)$$

$$\xi_E[n], \xi_I[n] \overset{\text{i.i.d.}}{\sim} \mathcal{N}(0,1) . \quad (41)$$

No $\sqrt{\Delta t}$ scaling was applied. These terms were therefore treated as discrete Gaussian input noise rather than as a formally defined continuous-time Gaussian white-noise process.

The equations were integrated using an explicit Euler scheme at 1 kHz, and both population activities were clipped to [0,1]. With $\Delta t = 0.001$ s,

$$E_{n+1} = clip_{[0,1]}\left[E_n + \frac{\Delta t}{\tau_E}\left(-E_n + S(u_{E,n})\right)\right], \tag{42}$$

$$I_{n+1} = clip_{[0,1]}\left[I_n + \frac{\Delta t}{\tau_I}\left(-I_n + S(u_{I,n})\right)\right], \tag{43}$$

where

$$u_{E,n} = w_{EE}E_n - w_{EI}I_n + I_{ext,n} + \eta_E[n], \quad u_{I,n} = w_{IE}E_n - w_{II}I_n + \eta_I[n]. \tag{44}$$

The initial conditions were $E_0 = 0.1$ and $I_0 = 0.05$. The independently searched back-end parameter set was

$$\Theta_{WC} = \{\tau_E, \tau_I, w_{EE}, w_{EI}, w_{IE}, w_{II}, \theta, \sigma, \sigma_\eta\}, \tag{45}$$

where $\sigma_\eta$ corresponds to *noise_std*. These quantities were interpreted as effective population-level dynamical parameters rather than direct estimates of individual synaptic, cellular, or receptor-level quantities. The excitatory output $E(t)$, implemented as *rE* in the code, was used to compute the model-derived $\gamma\%$, ITPC, and 40-Hz response amplitude.

For each simulated trial, the complete model mapping can be written as

$$s(t) \xrightarrow{F_{\mathrm{FE}}} I_{ext}(t) \xrightarrow{F_{\mathrm{WC}}} E(t), \tag{46}$$

where $F_{\mathrm{FE}}$ denotes the auditory front-end transformation and $F_{\mathrm{WC}}$ denotes the Wilson-Cowan back-end dynamics.

## Model fitting and configurations

The model was fitted using EEG-derived $\gamma\%$ and ITPC group means as empirical constraints. Let the model-generated values be

$$\hat{y} = (\hat{\gamma}_{HC}, \hat{\gamma}_{SCZ}, \hat{c}_{HC}, \hat{c}_{SCZ}), \tag{47}$$

and let the EEG-derived target vector be

$$y_{target} = (\gamma^*_{HC}, \gamma^*_{SCZ}, c^*_{HC}, c^*_{SCZ}) \ . \quad (48)$$

The three searches used experiment-specific normalized squared-error objectives rather than a single common absolute-error loss. For group $g$, the single-group objective was represented as

$$\mathcal{L}_g = a_\gamma \left(\frac{\hat{\gamma}_g - \gamma^*_g}{s_\gamma}\right)^2 + a_c \left(\frac{\hat{c}_g - c^*_g}{s_c}\right)^2 + P^{window}_g \ , \quad (49)$$

where $P^{\text{window}}_g$ denotes experiment-specific soft penalties for outputs outside the prescribed target region. Experiment 1 used $a_\gamma$=6.5, $s_\gamma$=1.0 , $a_c$=7.5 , and $s_c = 0.045$, together with group-specific guard penalties. Experiments 2 and 3 used $a_\gamma = a_c = 5.5$, $s_\gamma = 1.2$, and $s_c = 0.035$.

Define the model-generated group differences as

$$\Delta\hat{\gamma} = \hat{\gamma}_{HC} - \hat{\gamma}_{SCZ} \ , \quad \Delta\hat{c} = \hat{c}_{HC} - \hat{c}_{SCZ} \ , \quad (50)$$

with target differences

$$\Delta\gamma^* = 6.020 \ , \Delta c^* = 0.067 \ . \quad (51)$$

For Experiments 2 and 3, paired candidate solutions were ranked using

$$\mathcal{L}_{pair} = \mathcal{L}_{HC} + \mathcal{L}_{SCZ} + 2\left(\frac{\Delta\hat{\gamma} - \Delta\gamma^*}{2.0}\right)^2 + 5\left(\frac{\Delta\hat{c} - \Delta c^*}{0.045}\right)^2 + P_{direction} \ , \quad (52)$$

where $P_{\text{direction}}$ penalized failures to reach the required minimum HC-SCZ differences and, in Experiments 2 and 3, the required auxiliary $A_{40}$ direction. Experiment 1 used the sum of the two single-group losses together with additional penalties when the $\gamma$% or ITPC direction was reversed.

A final candidate was classified as accepted only when all experiment-specific hard criteria were satisfied. For all three experiments,

$$39.8 \le \hat{\gamma}_{\mathrm{HC}} \le 43.2,\ \ 33.8 \le \hat{\gamma}_{\mathrm{SCZ}} \le 37.2\ , \tag{53}$$

$$0.26 \le \hat{c}_{\mathrm{HC}} \le 0.40,\ \ 0.20 \le \hat{c}_{\mathrm{SCZ}} \le 0.32\ . \tag{54}$$

Experiment 1 additionally required

$$\Delta\hat{\gamma} \ge 3.5\ , \Delta\hat{c} \ge 0.015\ , \tag{55}$$

whereas Experiments 2 and 3 required

$$\Delta\hat{\gamma} \ge 3.0\ , \Delta\hat{c} \ge 0.015\ , \tag{56}$$

and

$$\Delta\widehat{A}_{40} = \widehat{A}_{40,HC} - \widehat{A}_{40,SCZ} \ge 0\ . \tag{57}$$

Amp40 was not used as a hard acceptance condition in Experiment 1.

The 40-Hz amplitude measure, denoted by $A_{40}$ , was used as an auxiliary direction index rather than as a primary EEG fitting target. It was computed from the 40-Hz narrowband envelope as a baseline-normalized response amplitude,

$$A_{40} = 10\ \log_{10}\!\left(\frac{\frac{1}{|T_{ana}|}\sum_{t\in T_{ana}} a_{40}^{2}(t)}{\frac{1}{|T_{base}|}\sum_{t\in T_{base}} a_{40}^{2}(t) + \varepsilon}\right)\ , \tag{58}$$

Where $a_{40}(t)$ is the Hilbert envelope after 38-42 Hz filtering, $T_{\mathrm{base}}$ is the baseline interval, $T_{\mathrm{ana}}$ is the post-onset analysis interval, and $\varepsilon$ is a small constant. In Experiment 1, $A_{40}$ was calculated directly from $E(t)$ without an additional observation-noise layer. In Experiments 2 and 3, fixed group-specific observation-noise standard deviations of $\sigma_{\mathrm{obs,HC}} = 0.07142$ and $\sigma_{\mathrm{obs,SCZ}} = 0.06734$ were used only for the auxiliary $A_{40}$ calculation, matching the corresponding search implementations. Observation noise was never applied to the model-derived $\gamma$% or ITPC. Consequently, $A_{40}$ was interpreted primarily as a within-experiment directional index rather than as a directly comparable fitted magnitude across the three configurations.

Three mechanistic parameter-search configurations were evaluated. In Experiment 1, referred to as the front-end-only model, HC and SCZ were allowed to have different auditory front-end parameters, while both groups shared the same fixed Wilson-Cowan back-end:

$$\Theta_{FE}^{HC} \neq \Theta_{FE}^{SCZ}, \Theta_{WC}^{HC} \neq \Theta_{WC}^{SCZ} = \Theta_{WC}^{0}. \quad (59)$$

This experiment tested whether differences in the searched auditory front-end parameters were sufficient to reproduce the HC-SCZ ASSR contrast while the Wilson-Cowan back-end was held fixed.

In Experiment 2, referred to as the back-end-only parameter search, the auditory front-end parameterization was fixed to the HC front-end obtained from Experiment 1, while HC and SCZ were allowed to have different Wilson-Cowan back-end parameters:

$$\Theta_{FE}^{HC} = \Theta_{FE}^{SCZ} = \Theta_{FE}^{HC,Exp1}, \Theta_{WC}^{HC} \neq \Theta_{WC}^{SCZ}. \quad (60)$$

This experiment tested whether differences in the searched Wilson-Cowan parameter block could reproduce the ASSR phenotype when both groups used the same fitted front-end parameterization. The labels "front-end only" and "back-end only" refer strictly to the parameter blocks that were independently searched. The fixed group-specific phase-dispersion settings defined above were retained in all experiments; consequently, Experiment 2 did not impose identical stochastic stimulus realizations across groups.

In Experiment 3, referred to as the full joint model, both the auditory front-end and the Wilson-Cowan back-end were allowed to vary between HC and SCZ:

$$\Theta^{HC}_{FE} \neq \Theta^{SCZ}_{FE}, \Theta^{HC}_{WC} \neq \Theta^{SCZ}_{WC}. \tag{61}$$

This configuration represented the complete front-end/back-end model and was used to evaluate joint changes in auditory input transformation and cortical E/I response properties.

For cross-model descriptive comparison, a common post-hoc absolute error score was calculated as

$$\varepsilon_{posthoc} = \left|\hat{\gamma}_{HC} - \gamma^{*}_{HC}\right| + \left|\hat{\gamma}_{SCZ} - \gamma^{*}_{SCZ}\right| + 10\left|\hat{c}_{HC} - c^{*}_{HC}\right| + 10\left|\hat{c}_{SCZ} - c^{*}_{SCZ}\right| . \tag{62}$$

This score was used only to summarize numerical deviations from the EEG targets across the final models and in sensitivity reporting. It was not the optimization loss used to obtain the accepted solutions.

## Robustness and dynamical analyses

Only the parameters designated as free in each experiment were perturbed. In Experiment 1, the HC and SCZ auditory front-end parameters were perturbed while the shared Wilson-Cowan back-end remained fixed. In Experiment 2, only the HC and SCZ back-end parameters were perturbed while the common auditory front-end parameterization remained fixed. In Experiment 3, both groups' front-end and back-end parameters were perturbed.

For a fitted parameter vector $\boldsymbol{\theta}^{*}$ and perturbation level $\rho$, each free parameter was perturbed as

$$\theta^{(m,\rho)}_{i} = clip_{[l_i,u_i]}\,[\theta^{*}_{i}\,(1+\xi^{(m)}_{i})], \qquad \xi^{(m)}_{i} \sim \mathcal{U}(-\rho\,,\rho)\,, \tag{63}$$

where $m = 1,\dots,M$ indexes perturbation repeats, $\rho \in \{0.05, 0.10\}$, and $[l_i,u_i]$ is the predefined admissible search range of parameter $i$. Thus, the perturbations were bounded multiplicative perturbations rather than unconstrained symmetric changes.

Front-end perturbations that violated the Nyquist-feasibility condition in Eq.24 were rejected and redrawn. For each perturbation level, $M = 100$ random parameter vectors were evaluated. The formal sensitivity analysis used $N_{\text{trials}} = 60$ and $n_{\text{ch}} = 128$, and the unperturbed baseline case $\rho = 0$ was required to reproduce the accepted model results within strict numerical tolerances. Independent Monte Carlo perturbation vectors were drawn for the two perturbation levels.

The primary perturbation sensitivity measure was the joint preservation rate of the $\gamma$% and ITPC group directions:

$$R_{\gamma+c}(\rho) = \frac{1}{M} \sum_{m=1}^{M} \mathbf{1}\,[\hat{\gamma}_{HC}^{(m,\rho)} > \hat{\gamma}_{SCZ}^{(m,\rho)}]\,\mathbf{1}\,[\hat{c}_{HC}^{(m,\rho)} > \hat{c}_{SCZ}^{(m,\rho)}]\,. \tag{64}$$

An additional three-index direction rate was computed by including the auxiliary 40-Hz amplitude direction:

$$R_{\gamma+c+A}(\rho) = \frac{1}{M} \sum_{m=1}^{M} \mathbf{1}\,[\hat{\gamma}_{HC}^{(m,\rho)} > \hat{\gamma}_{SCZ}^{(m,\rho)}]\,\mathbf{1}\,[\hat{c}_{HC}^{(m,\rho)} > \hat{c}_{SCZ}^{(m,\rho)}]\,\mathbf{1}\,[\hat{A}_{40,HC}^{(m,\rho)} > \hat{A}_{40,SCZ}^{(m,\rho)}]\,. \tag{65}$$

The target-window pass rate was also reported as a secondary index. Because requiring randomly perturbed models to remain inside all target windows is stricter than preserving the empirical HC-SCZ direction, the target-window rate was not used as the primary robustness criterion.

For dynamical analysis, the accepted parameters were fixed and no additional parameter search was performed. The front-end-derived operating input was estimated using 60 independently generated auditory trials and the full 128-channel front-end per condition, matching the resolution used in the final model evaluations. The fixed-point and Jacobian calculations used the accepted Wilson-Cowan parameter sets without modification.

Let $\mathcal{O}$ denote the samples from 0.25 s after stimulus onset to the end of the 2.0-s stimulus. The operating input was defined as

$$I_0^{op} = \frac{1}{K_{op}} \sum_{k=1}^{K_{op}} [\frac{1}{|O|} \sum_{n \epsilon O} I_{ext,k}[n]] , K_{op} = 60 . \quad (66)$$

The trial-averaged 5th and 95th percentiles of $I_{\text{ext}}$ over the same interval were used to represent the operating range.

The Wilson-Cowan back-end was then analyzed under constant external input $I_0$ with noise removed. The deterministic system was

$$\tau_E \frac{dE}{dt} = -E + S(w_{EE}E - w_{EI}I + I_0) , \quad (67)$$

$$\tau_I \frac{dI}{dt} = -I + S(w_{IE}E - w_{II}I) . \quad (68)$$

For each constant input value $I_0$, fixed points $(E^*, I^*)$ satisfy

$$E^* = S(w_{EE}E^* - w_{EI}I^* + I_0) , \quad (69)$$

$$I^* = S(w_{IE}E^* - w_{II}I^*) . \quad (70)$$

Multiple initial guesses were used at each input value to identify coexisting numerical fixed points. For the input scan, the root solver was initialized from a 6×6 grid over $[0.02,0.98]^2$, together with additional low-, high-, and central-activity initial guesses. At the front-end-derived operating input, at least an 8×8 initialization grid was used. Roots outside $[0,1]^2$, roots with a residual norm greater than $10^{-6}$, and numerical duplicates separated by less than $10^{-5}$ were discarded.

When several fixed points were detected at the operating input, locally stable solutions were considered first. The primary operating-point solution was the fixed point whose excitatory coordinate was closest to the mean excitatory activity of the corresponding periodically forced deterministic simulation:

$$(E_{op}^*, I_{op}^*) = \underset{(E^*, I^*) \in c}{\arg\min} \left| E^* - \overline{E}_{forced} \right| , \quad (71)$$

where

$$C = \begin{cases} S_{stable}, & S_{stable} \neq \varnothing \\ S_{all}, & S_{stable} \neq \varnothing \end{cases} . \quad (72)$$

Thus, if one or more stable fixed points were available, the closest stable solution was selected; otherwise, the closest available fixed point was selected.

The local stability of each fixed point was determined by linearizing the system around ($E^*$ , $I^*$ ). Let

$$u_E^* = w_{EE}E^* - w_{EI}I^* + I_0 , u_I^* = w_{IE}E^* - w_{II}I^* . \quad (73)$$

The derivative of the sigmoid is

$$S'(u) = \frac{S(u)[1-S(u)]}{\sigma} . \quad (74)$$

Thus, the Jacobian matrix at the fixed point is

$$J(E^* , I^* ; I_0) = \begin{pmatrix} \frac{-1 + w_{EE} S'(u_E^*)}{\tau_E} & \frac{-w_{EI} S'(u_E^*)}{\tau_E} \\ \frac{w_{IE} S'(u_I^*)}{\tau_I} & \frac{-1 - w_{II} S'(u_I^*)}{\tau_I} \end{pmatrix} . \quad (75)$$

Let $\lambda_1$ and $\lambda_2$ be the eigenvalues of $J$. The maximum real part was defined as

$$\lambda_{max} = max \{Re(\lambda_1),Re(\lambda_2)\} . \quad (76)$$

A fixed point was classified as locally asymptotically stable if

$$\lambda_{max} < 0 , \quad (77)$$

and locally unstable under the constant-input approximation if

$$\lambda_{max} > 0 . \quad (78)$$

Only values of $\lambda_{max}$ sufficiently close to zero were described as proximity to a local stability boundary.

The trace, determinant, and discriminant of the Jacobian were

$$\mathrm{tr}\,(J) = J_{11} + J_{22}\ ,\mathrm{det}\,(J) = J_{11}J_{22} - J_{12}J_{21}\ , \tag{79}$$

$$\Delta_J = \mathrm{tr}\,(J)^2 - 4\mathrm{det}\,(J)\ . \tag{80}$$

For a two-dimensional continuous-time system, $\det(J) < 0$ indicates a saddle point. When $\det(J) > 0$, a negative trace indicates local stability and a positive trace indicates local instability. The condition $\Delta_J < 0$ indicates a complex-conjugate eigenvalue pair.

Points with $\det(J) > 0$ and small $|tr\,(J)|$ were treated as trace-zero proximity indicators. The screening procedure did not additionally require $\Delta_J < 0$ and therefore did not verify that the relevant eigenvalues formed a complex-conjugate pair. Because the analysis did not perform numerical continuation, test transversality, or calculate normal-form coefficients, these points were not classified as formal Hopf bifurcations. Similarly, small values of $|\det(J)|$ were described as saddle-node-like proximity rather than as proof of a saddle-node bifurcation.

The constant input was sampled at 161 equally spaced values,

$$I_{0,k} = \frac{1.2k}{160}\ , \quad k = 0,\ldots,160\ , \tag{81}$$

corresponding to

$$I_0 \in [0,1.2]\ . \tag{82}$$

The resulting fixed points and local-stability labels were used to construct input-driven fixed-point diagrams. The mean front-end-derived operating input and its 5th-95th percentile range were overlaid on these diagrams to relate the driven operating regime to the local back-end dynamical landscape.

# Results

## Empirical EEG contrasts used as model constraints

The final EEG subset included 21 HC and 21 SCZ participants. The empirical group means were higher in HC than in SCZ for both primary measures, although substantial inter-individual variability was present (Table 1). For $\gamma$%, the HC mean was 41.496 (SD = 26.668), compared with 35.476 (SD = 27.399) in SCZ, corresponding to an HC-SCZ mean difference of 6.020 percentage points. The 95% bootstrap confidence interval for this difference ranged from -9.442 to 21.757. The Welch comparison was not statistically significant ($t_{39.97} = 0.722$, P=.475; Holm-adjusted $P = .475$), with a small standardized effect size (Hedges' $g = 0.218$). The Mann-Whitney sensitivity analysis was also nonsignificant ($P = .450$).

For ITPC, the HC group had a mean value of 0.325 (SD = 0.153), whereas the SCZ group had a mean value of 0.258 (SD = 0.103). The corresponding HC-SCZ difference was 0.0667, with a 95% bootstrap confidence interval from -0.0109 to 0.1441. The Welch test yielded $t_{35.05}$=1.658 and P=.106 (Holm-adjusted $P = .212$), with Hedges' $g = 0.502$. The Mann-Whitney test was likewise nonsignificant ($P = .227$).

Thus, the empirical group means showed the expected *HC>SCZ* direction for both $\gamma$% and ITPC, but the present participant-level sample did not establish statistically significant group differences. Accordingly, these values were used as directional group-level numerical constraints for model fitting rather than as evidence of statistically significant deficits in this particular subset.

## Overall model performance

Table 2 summarizes the EEG-derived targets and final outputs of the three primary model configurations. All three configurations satisfied the prespecified fitting criteria and reproduced higher γ% and ITPC values in HC than in SCZ. The front-end-restricted, back-end-restricted, and full-joint models yielded HC–SCZ γ% differences of 6.248, 6.394, and 6.302 percentage points, respectively, compared with the empirical difference of 6.020. The corresponding ITPC differences were 0.042, 0.069, and 0.073, compared with the empirical difference of 0.067. Thus, similar group-level EEG constraints were reproduced despite different restrictions on the front-end and back-end parameter blocks. The Experiment 1 gain-matched analysis was retained only as an auxiliary control and was not treated as a fourth mechanistic model.

## Front-end-restricted model

Experiment 1 tested whether the HC–SCZ contrast could be reproduced with a shared Wilson–Cowan back-end while fitting the auditory front-end separately for the two groups. The accepted solution yielded γ% values of 41.625 and 35.377 and ITPC values of 0.297 and 0.255 for HC and SCZ, respectively, preserving the HC > SCZ direction for both primary measures. The auxiliary 40-Hz amplitude was also higher in HC than in SCZ (14.959 vs 9.352 dB).

The fitted front-ends differed across multiple effective signal-processing parameters. In particular, the SCZ solution combined a lower compression exponent with a substantially longer temporal-integration constant. Both fitted front-ends satisfied the imposed digital-filter feasibility criterion, with their highest filter edges remaining below the 7990-Hz numerical limit.

Because the Wilson–Cowan back-end was identical between groups, this experiment demonstrates that group-specific back-end parameters were not required to reproduce the observed HC > SCZ direction within the tested model family.

## Back-end-restricted model

Experiment 2 used the same fitted HC auditory front-end for both groups while allowing the Wilson–Cowan back-end parameters to differ. The accepted model yielded an HC–SCZ γ% difference of 6.394 percentage points and an ITPC difference of 0.069, closely matching the empirical group differences. The auxiliary 40-Hz amplitude was also higher in HC than in SCZ (5.023 vs 2.736 dB).

The fitted back-ends differed in effective time constants, coupling strengths, sigmoid slope, and discrete input-noise parameters. These parameter sets define different effective Wilson–Cowan dynamical regimes and should not be interpreted as direct estimates of specific receptors, interneurons, or synaptic connections.

## Full joint front-end/back-end model

Experiment 3 allowed both the auditory front-end and Wilson–Cowan back-end to vary between HC and SCZ. The accepted solution yielded γ% values of 41.534 and 35.232 and ITPC values of 0.319 and 0.245, respectively, corresponding to HC–SCZ differences of 6.302 percentage points and 0.073. The auxiliary 40-Hz amplitude was also higher in HC (5.354 dB) than in SCZ (1.905 dB).

The accepted SCZ front-end showed a lower compression exponent and a longer temporal-integration constant than the HC front-end, while the fitted Wilson–Cowan parameter sets also differed between groups. Both front-ends satisfied the Nyquist-feasibility criterion. Thus, the full-joint solution distributed the fitted HC–SCZ contrast across both model stages, although its close numerical fit does not establish that this configuration is uniquely correct.

## Local parameter sensitivity

The unperturbed sensitivity baselines reproduced the accepted model solutions before parameter perturbation. For each model, 100 random parameter vectors were evaluated at each perturbation level (±5% and ±10%). The primary criterion was simultaneous preservation of the HC > SCZ directions for γ% and ITPC.

The joint γ%–ITPC direction-preservation rates were 56% and 70% for the front-end-restricted model, 88% and 74% for the back-end-restricted model, and 96% and 86% for the full-joint model at ±5% and ±10%, respectively (Figure 1). The apparent increase for Experiment 1 at the larger perturbation level does not indicate greater robustness because independent Monte Carlo draws were used at the two levels. Exact target-window retention was substantially lower, indicating that direction preservation was more robust than precise numerical agreement with all fitting windows. These analyses characterize only local neighborhoods around the selected solutions and do not establish global identifiability or formal superiority of the full-joint model.

## Fixed-point structure and local stability

Deterministic fixed-point analysis was performed on the accepted Wilson–Cowan parameter sets using front-end-derived operating inputs. In Experiment 1, both HC and SCZ operating points were locally stable within the common back-end landscape. In Experiment 2, the HC operating-point fixed point was locally unstable under the constant-input approximation, whereas the SCZ operating point was locally stable. The HC back-end also approached a trace-zero condition, while the SCZ back-end exhibited up to three numerical fixed points over part of the scanned input range. Because numerical continuation was not performed, these observations were not assigned formal Hopf or saddle-node bifurcation classifications.

In Experiment 3, both group-specific operating points were locally stable. Thus, similar ASSR outputs were compatible with different local dynamical organizations across the fitted model configurations (Figure 2). These results characterize the effective model dynamics and should not be interpreted as direct statements about physiological stability in HC or SCZ participants.

## Discussion

This study examined whether schizophrenia-related 40-Hz ASSR abnormalities could be reproduced by model variation at different levels of an auditory-processing framework. Although HC means exceeded SCZ means for both γ% and ITPC, the participant-level group differences were not statistically significant; the empirical values were therefore used primarily as group-level numerical constraints. Under these constraints, the front-end-restricted, back-end-restricted, and full-joint models all satisfied the prespecified fitting criteria and reproduced the HC > SCZ direction.

The three configurations provided different candidate mechanistic accounts. The front-end-restricted model showed that changes in effective auditory sensory encoding and input transformation could reproduce the ASSR pattern while both groups shared the same Wilson–Cowan back-end. These front-end processes represent effective auditory filtering, nonlinear transduction, and temporal integration[22,23]. In contrast, the back-end-restricted model showed that altered cortical excitatory/inhibitory (E/I) population dynamics could reproduce a similar phenotype under a common auditory front-end, consistent with previous work linking schizophrenia-related gamma abnormalities to cortical E/I function[3,13-15,18,20,21]. The full-joint model further showed that combined changes in auditory sensory encoding and cortical dynamics could also reproduce the same group-level pattern.

An important implication is therefore that schizophrenia-related 40-Hz ASSR abnormalities may arise through different candidate mechanistic pathways. In some patients, the electrophysiological abnormality may be more consistent with altered auditory sensory encoding or effective input transformation; in others, it may be more consistent with altered cortical E/I dynamics; and in still others, both stages may contribute. Thus, patients with similar macroscopic ASSR abnormalities may nevertheless differ in their underlying model-level mechanisms. This provides a computational account of potential mechanistic heterogeneity in schizophrenia-related electrophysiology. These findings concern candidate mechanisms underlying the ASSR abnormality and do not establish that these changes directly cause schizophrenia itself.

The sensitivity and fixed-point analyses further supported differences among the accepted models. The full-joint model showed the highest tested joint γ%–ITPC direction-preservation rates within the examined local parameter ranges, although this does not establish superiority over the other configurations[28]. Different local dynamical organizations were also compatible with similar ASSR outputs, indicating that comparable electrophysiological phenotypes need not reflect identical internal dynamics. Because formal numerical continuation and bifurcation analyses were not performed, these dynamical results should be viewed as complementary descriptions of the fitted models.

The three-model framework also suggests a potential direction for future participant-level analysis. Generative-modeling studies in computational psychiatry have shown the potential of mechanistic models for individual-level inference and patient stratification[29-31]. Future studies could fit all three configurations to individual patients and compare their relative fitting or predictive performance to assess whether an electrophysiological profile is more consistent with altered sensory encoding, altered cortical E/I dynamics, or combined changes. Following validation with independent patient data and clinical studies, this framework may help support individualized mechanistic assessment and clinical stratification.

Several limitations should be considered. The empirical sample was modest and the participant-level group differences were not statistically significant. EEG channel averaging reduced spatial information[4,27], empirical ITPC was not strictly stimulus-onset locked, and the empirical and simulated ASSR protocols were not identical[24]. In addition, both the auditory front-end and the Wilson–Cowan back-end are simplified effective models, and γ% and ITPC are summary measures for which different parameter combinations may remain observationally equivalent[28].

In conclusion, similar schizophrenia-related 40-Hz ASSR abnormalities can be reproduced by changes in auditory sensory encoding, cortical E/I dynamics, or both. The three configurations therefore provide a computational framework for understanding potential mechanistic heterogeneity across patients and, following future clinical validation, may support individualized mechanistic assessment and clinical stratification.

doi:10.1016/j.nicl.2013.11.002.

# Tables

**Table 1**: Participant-level EEG summary and HC–SCZ group comparisons.

| Measure | HC mean(SD) | SCZ mean(SD) | Mean differences(95% CI) | Welch *t*(df), | *P* | Holm *P* | Hedges' *g* | MWU *P* |
|---|---|---|---|---|---|---|---|---|
| γ% | 41.496(26.668) | 35.476(27.399) | 6.020(-9.442,21.757) | 0.722(39.97), | 0.475 | 0.475 | 0.218 | 0.450 |
| ITPC | 0.325(0.153) | 0.258(0.103) | 0.0667(-0.0109,0.1441) | 1.658(35.05), | 0.106 | 0.212 | 0.502 | 0.227 |

**Note.** HC, healthy controls; SCZ, participants with schizophrenia; SD, standard deviation; γ%, gamma-band suprathreshold proportion; ITPC, inter-trial phase consistency; CI, confidence interval; MWU, two-sided Mann–Whitney U sensitivity test. Confidence intervals represent 10,000-resample percentile bootstrap confidence intervals for the HC–SCZ mean difference. Holm adjustment was applied to the two Welch tests.

**Table 2:** EEG-derived targets and final outputs of the model configurations.

| Model | $\gamma\%_{HC}$ | $\gamma\%_{SCZ}$ | Δγ | $ITPC_{HC}$ | $ITPC_{SCZ}$ | ΔITPC | Accepted |
|---|---|---|---|---|---|---|---|
| EEG-derived target | 41.496 | 35.476 | 6.020 | 0.325 | 0.258 | 0.067 | - |
| Experiment 1:front − end restricted | 41.625 | 35.377 | 6.248 | 0.297 | 0.255 | 0.042 | Yes |
| Experiment 2:back − end restricted | 42.227 | 35.833 | 6.394 | 0.319 | 0.250 | 0.069 | Yes |
| Experiment 3:full joint | 41.534 | 35.232 | 6.302 | 0.319 | 0.245 | 0.073 | Yes |
| Experiment1:gain − matched control | 41.625 | 35.421 | 6.204 | 0.297 | 0.263 | 0.034 | Yes |

**Note.** HC, healthy controls; SCZ, participants with schizophrenia; γ%, gamma-band suprathreshold proportion; ITPC, inter-trial phase consistency. Δγ = γ%HC − γ%SCZ, and ΔITPC = ITPCHC − ITPCSCZ. "Accepted" indicates that the corresponding model configuration satisfied the prespecified fitting criteria. Differences were calculated from unrounded simulation outputs; therefore, a reported difference may differ by 0.001 from direct subtraction of the displayed rounded values. The Experiment 1 gain-matched control is an auxiliary analysis and is not treated as a fourth mechanistic model.

# Figure legends

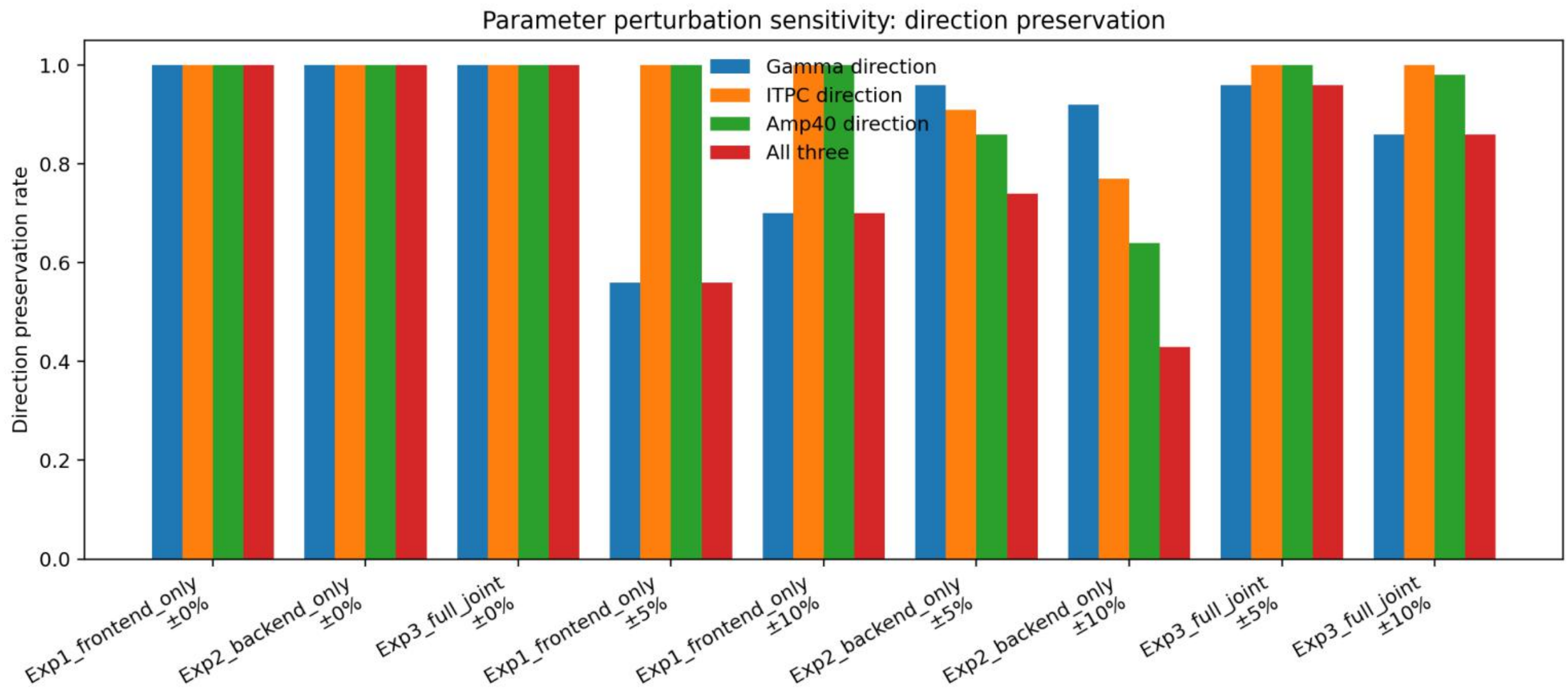

**Figure 1. Direction preservation under the final local parameter-perturbation analysis.** For each model configuration, 100 randomly perturbed parameter vectors were evaluated at each perturbation level (±5% and ±10%) using 60 trials and the complete 128-channel auditory front-end. Bars show the proportions of simulations preserving the HC > SCZ direction for the gamma-band suprathreshold proportion (γ%), inter-trial phase consistency (ITPC), and auxiliary 40-Hz amplitude (A40), together with simultaneous preservation of all three directions. The primary robustness criterion reported in the text was simultaneous preservation of the γ% and ITPC directions. The ±0% entries indicate the unperturbed accepted baselines.The full-joint configuration showed the highest tested joint γ%–ITPC direction-preservation rates at both perturbation levels. Because independent Monte Carlo perturbation draws were used at the two perturbation levels, the rates should not be interpreted as a monotonic function of perturbation magnitude.

**Alt text:** Grouped bar chart comparing direction-preservation rates under ±5% and ±10% parameter perturbations for the front-end-restricted, back-end-restricted, and full-joint models. Bars show preservation of the HC > SCZ direction for γ%, ITPC, A40, and all three measures simultaneously. The full-joint model shows the highest tested joint γ%–ITPC preservation at both perturbation levels.

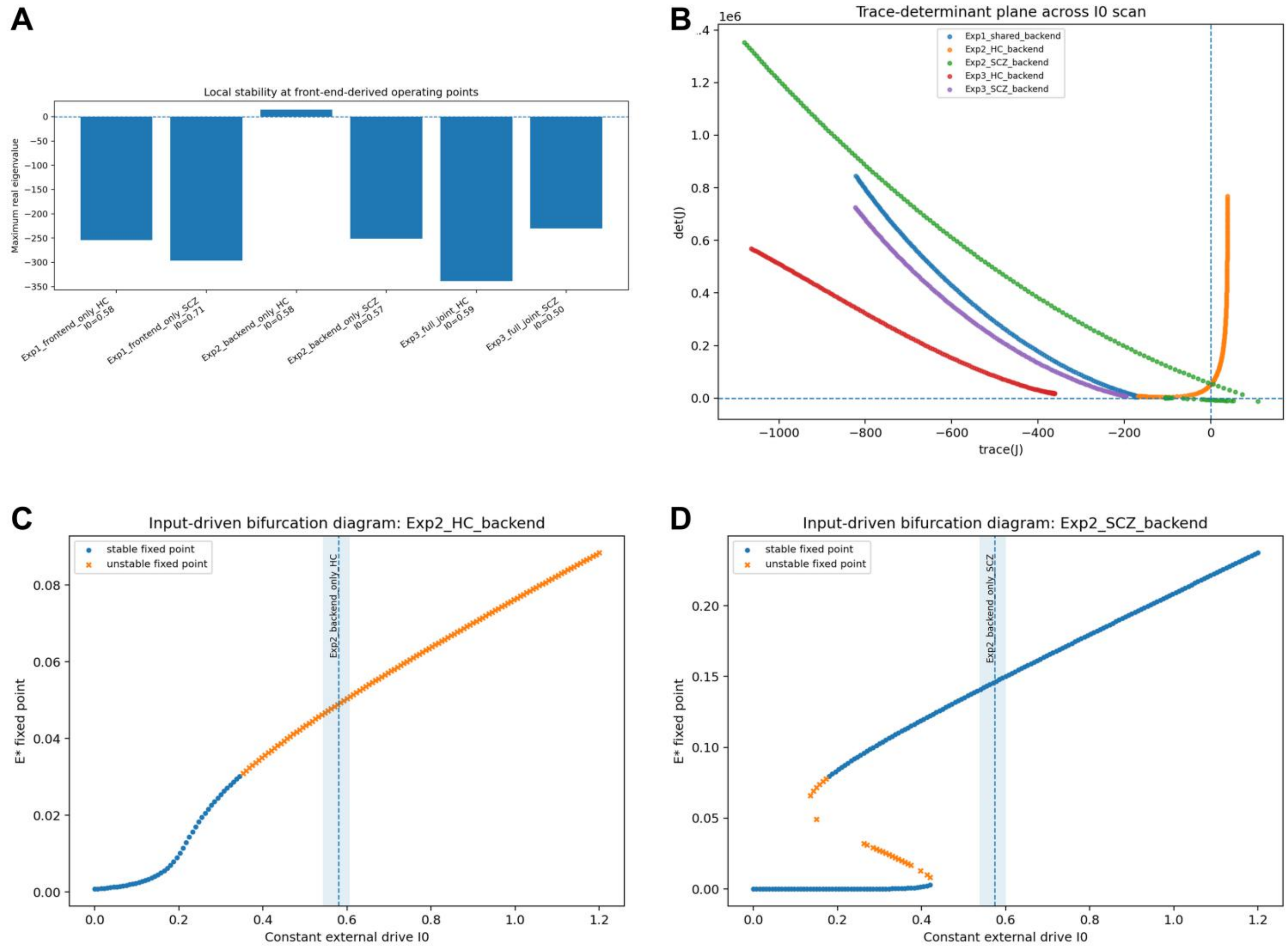

**Figure 2. Post-hoc deterministic dynamical analysis of the final fitted Wilson–Cowan back-ends. A,** Maximum real part of the Jacobian eigenvalues at each front-end-derived operating point. Negative values indicate local asymptotic stability under the deterministic constant-input approximation; the Experiment 2 HC operating point was the only fitted condition with a positive maximum real eigenvalue. **B,** Trace–determinant trajectories of the fitted back-ends across the constant-input scan. The Experiment 2 HC back-end approached and crossed the trace-zero boundary; these observations were treated only as local stability-boundary proximity indicators and were not classified as formal Hopf bifurcations. **C,** Input-driven fixed-point diagram for the Experiment 2 HC back-end. The front-end-derived HC operating input was located on a locally unstable portion of the numerical fixed-point branch. **D,** Input-driven fixed-point diagram for the Experiment 2 SCZ back-end. Multiple numerical fixed points were identified over part of the scanned input range, indicating numerical multistability, whereas the final SCZ operating point was locally stable. These analyses characterize the fitted effective models under a constant-input approximation and should not be interpreted as direct statements about physiological stability in HC or SCZ participants.

**Alt text:** Four-panel figure summarizing deterministic Wilson–Cowan dynamics. Panel A compares the maximum real Jacobian eigenvalue across fitted operating points and shows only the Experiment 2 HC condition above zero. Panel B shows trace–determinant trajectories across the constant-input scan. Panel C shows the Experiment 2 HC fixed-point branch with the fitted operating point on a locally unstable branch. Panel D shows the Experiment 2 SCZ fixed-point structure, including multiple numerical fixed points over part of the input range and a locally stable fitted operating point.